**Title:** Using Information Semantic Systems for Absolutely Secure Processing.
**Authors:** Vadim Nedzvetski

Propose a new cryptographic information concept. It allows :

- to create absolutely algorithmic unbreakable ciphers for communication through open digital channels;
- to create new code-breaking methods. They will be the most efficient decoding methods to-date, with the help of which any of the existing codes could, in principle, be broken, provided it is not absolutely unbreakable.

## C o n t e n t s.

# 1. Algorithmic Theory of Information Applied to Cryptology.

## 1.1. Disadvantages of the Shannon's Theory of Information.

The word "information" is used in different senses. In common life, in computer science, in the statistical, or in the theory of communication we use this word in different ways depending on its applications. It is commonly thought that cryptology is the science of secure communications [10] and then the theory of communication must be laid as a theoretical base of cryptology. For this reason modern cryptology uses its idea and notions that were stated by C. Shannon in his classical papers [8, 9]. But there are three reasons that force us to look for another information theoretical base for cryptology :

**1.** There is a difference between the Shannon's theory approach and cryptology needs :

The theory of communication studies problems of transmitting information from one correspondent to another. It is not interested in the content of the message. In this reason the theory uses probabilistic definition of information. Cryptology concerns with the actual message content.

**2.** Shannon's theory does not treat the semantic content of a message. (It is not interested in the meaning of a message). It is interested a number of characters only. For that reason today's cryptography does not have methods for ciphering and deciphering of semantic information.

**3.** Communication theory cannot explain the public-key cryptographic concept.

According to the Shannon's theory amount of information **I(M; C)** of the message **M**, that can be revealed if a cipher **C** was intercepted is

$$\mathbf{I(M; C) = H(M) - H(M/C)}, \qquad (1.1)$$

where **H(M)** is amount of information (entropy) in the message and **H(M/C)** is the conditional amount of information (entropy) of the message when the encrypted text **C** (cipher) is available. But as far as for any public-key method entropy

$$\mathbf{H(M/C) = 0}, \qquad (1.2)$$

then any public-key method (according to Shannon's theory) is not really a cryptographic method because it does not hide any information. For any public-key method always

$$\mathbf{I(M; C) = H(M)}. \qquad (1.3)$$

To find the theoretical base for cryptology consider existing approaches to information.

## 1.2. Different approaches to the meaning of information.

There are 3 different mathematical approaches to meaning of information :

1. **Experimental approach.** R. A. Fisher (1925) [2]) thought about information as something what can be obtained from processing of experimental data.
2. **The Theory of Communication.** R. V. Hartley (1928) [3] defined information as a choice when a event was selected from a set of possible events (we will call it as information set). C. E. Shannon (1948) [8] defined information as a choice of the actual message from a set of possible messages.
3. **The Algorithmic Theory of Information.** A. A. Kolmogorov (1965) [4, 5] (see also [6]) was the first who introduced algorithmic ideas to approach of information. He defined information as algorithmic complexity of the object. (The algorithmic theory of information is not widely known. Articles [4]-[6] are not translated in English. In English literature only [1] is available.)

We see that there are different ideas about what is information :

- The experimental approach thinks about information as a property of the object that must be determined.
- The theory of communications defined information as a process of choice.
- The algorithmic theory considers information as an algorithm needed to rebuild the object.

The algorithmic theory is concerned with content of the actual message. For this reason it is more suitable for cryptographic needs. However, its authors did not consider cryptographic tasks. And today it is necessary to develop the algorithmic idea for cryptographic needs. To understand what we need consider a question about what is information in the most general case. Intuitively we imply that

***information about an object or event is any property or characteristic of the object that can be considered, transmitted or processed without necessity for physical presence of the object or event.*** **(1.4)**

In other words, information about an object is anything what can be considered separately from this object. This definition is not a mathematically strong one. So we should discuss it more detailed :

1. The abstraction from physical content of the object means that there is congruity between the object and a symbolic set or that we can make a mathematical or a symbolic model (representation) of the object.
2. The model of the object is built according to a research technique (algorithm) and experimental devices.
3. The model of the object itself are not needed for us. We use them to make our decision. This decision we make according to a processing algorithm. Symbolic representation of the object (its model) is input data for the processing algorithm. Our decision (choice) is output of the processing algorithm.
4. Any transfer from a real object to its model and from the model to a choice (information set) must be (if we want to be within the bounds of strong mathematical definitions) described by mathematical algorithms (research algorithm and processing algorithm).

5. A real object is a carrier of information. A choice from an information set is of real usefulness to us. But the model of the object is actual information about the object.

In the future we will call for our approach as the cryptographic approach.

Figure 1 shows the general schema of information and relationship between different approaches about the question what is information.

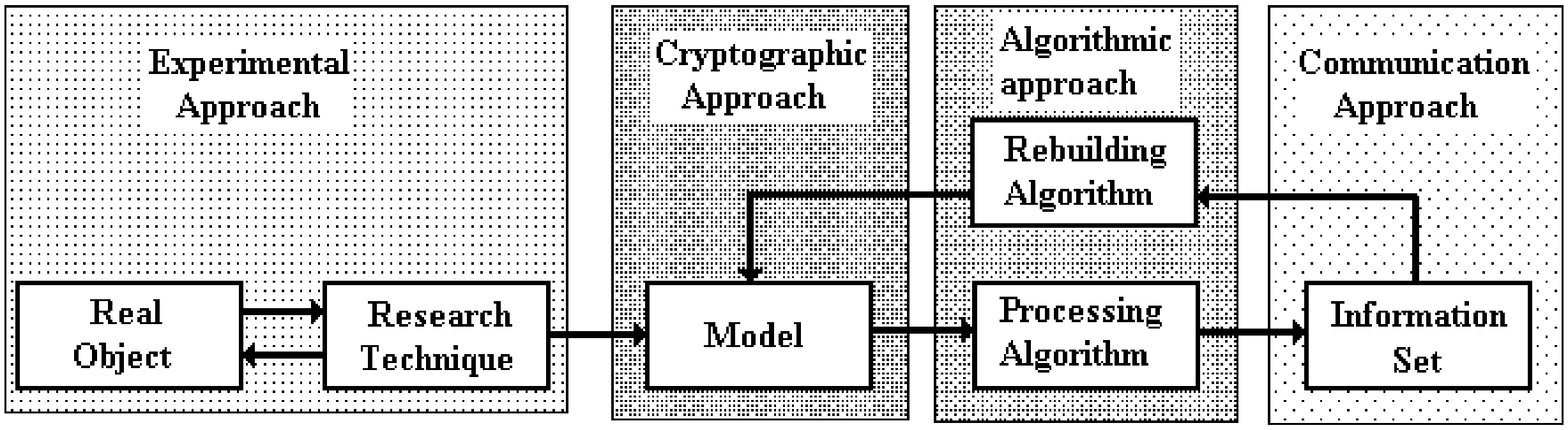

**Fig. 1. The general schema of obtaining information.**

There are different starting points for definition of information:

- an information set.
- a model.

**The communication theory** chooses a starting point for definition of information as an information set of possible properties or messages. **The algorithmic theory** supposes that information is algorithmic links between the model and its information set. What is difference ?

Definition **(1.4)** defined information as any property or characteristic of the object. The property of the object is one from the set of any possible properties. So if we have determined a property we have made a choice from a set of possible properties. However the real information is a model but we are interested in a choice. For this reason we omit often the previous stages : model and processing algorithm and think that real information is a choice from information set. But it will be right if

***for any predicate about an arbitrary model***
***an algorithm solving this predicate exists.*** **(1.5)**

C. Shannon published his papers in 1948-49 and wrote them early. In that time the presumption **(1.5)** seemed obvious. Hovewer in the following years many unsolvable models were found (The first algorithmic unsolvable mathematical models were built by A.A.Markov and E. L. Post in 1947). For these models the statement **(1.5)** is wrong. It means that :

1. For definition of information is necessary to take into consideration the research and processing algorithms that are used for obtaining information.

2. If we transfer from a object model to its information set we can lose information about the object model.

### 1.3. Semantic Information.

What information about the object can be lost if we transfer from its model to the its information set ? We saw above that this is information about the meaning of the predicate. So it is semantic information. Semantic information about the predicate expresses a link between elements of the model. This link we will call as a semantic relation. The set of elements with semantic relation we will call as semantic set. An algorithm that makes semantic relation we will call as a semantic algorithm. As opposed to semantic information Shannon's information we will call as coded information. Certainly there are a lot of links between elements expressed by algorithm can be shown as coded information. In order to avoid confusions in the future we will call semantic information only links that cannot be shown as coded information therefore these links cannot be algorithmic solved. So

***Semantic information about an object is algorithmic unsolvable links between elements of the object.*** **(1.6)**

Semantic information is expressed by algorithmic unsolvable problems. There are infinitely enumerable set of unsolvable problems that cannot be reduced to each other. Any of these problems can be used for formulation of semantic information. The properties of algorithmic unsolvable problems establish the properties of information systems for processing of semantic information.

Note the most important features of semantic information :

1. Semantic information cannot be revealed by any algorithm.
2. Semantic information may "go through" algorithms untouched.
3. There are an infinite number of forms for expression of the same semantic information.

The features 1 and 2 are important for cryptography and secure information system. It means that we can make absolutely secure communication and processing information systems. Any coded information may be expressed by links between elements. It means that we can use semantic information for transmitting and processing coded information.

Semantic information are properties of the object that cannot be reduced to coded information form. We called these properties as semantic properties or semantic information. But in order to these properties will be real information according to definition **(1.4)**, we should have possibility to transmit and process semantic information immediately by semantic form.

### 1.4. Transmitting of semantic information.

According to definition of semantic information, there is no algorithmic way to reveal a semantic relationship. Therefore, if we have transmitted a few symbols combined by semantic relationship, then our correspondent are not able to reveal the received semantic information. It is possible to transmit both symbols and semantic algorithm, but it will mean that we expressed semantic information by coded form. For applications it is important to answer the following question : can semantic information be transmitted immediately by semantic form ?

For transmitting semantic information in semantic form we propose the following protocol. Assume there are two correspondents Alice and Bob. Alice wishes to send semantic information to Bob. The correspondents do the following operations :

1. Alice makes a semantic relation the following way. Any semantic relation includes a set of objects and semantic algorithm that introduces a semantic relationship. Assume the set of objects is **M**. Alice's semantic algorithm is $\hat{\mathbf{A}}$. So $\hat{\mathbf{A}}\otimes\mathbf{M}$ is a new set with semantic relationship.
2. Alice transmits the set

$$\mathbf{C_1 = \hat{A}\otimes M} \quad \mathbf{(1.7)}$$

to Bob.

3. Bob cannot reveal the Alice's semantic relation. But he can make a conjecture about it. According to his conjecture he adds his semantic set **L**. As far as his set depend on Alice's message we write it symbolically as $(\hat{\mathbf{A}}\otimes\mathbf{M}\oplus\mathbf{L})$.
4. Bob uses his semantic relation $\hat{\mathbf{B}}$ and sends the result

$$\mathbf{C_2 = \hat{B}\otimes(\hat{A}\otimes M \oplus L)} \quad \mathbf{(1.8)}$$

back to Alice.

5. Alice is not able to reveal the Bob's semantic relation. But she knows her semantic set **M** and her semantic algorithm $\hat{\mathbf{A}}$. Alice can use the inverse algorithm $\hat{\mathbf{A}}^{-1}$ :

$$\mathbf{\hat{A}^{-1}\otimes C_2 = \hat{A}^{-1}\otimes\hat{B}\otimes(\hat{A}\otimes M \oplus L)} \quad \mathbf{(1.9)}$$

6. If **A** and **B** are commutative each other then

$$\mathbf{\hat{A}^{-1}\otimes\hat{B}\otimes(\hat{A}\otimes M \oplus L) = \hat{B}\otimes M \oplus \hat{A}^{-1}\otimes\hat{B}\otimes L} \quad \mathbf{(1.10)}$$

7. If the semantic set **M** has a special property then Alice uses it to separate the first and the second terms in **(1.10)**. In that case she will get

$$\mathbf{C_3 = \hat{B}\otimes M} \quad \mathbf{(1.11)}$$

$$\mathbf{C_4 = \hat{A}^{-1}\otimes\hat{B}\otimes L} \quad \mathbf{(1.12)}$$

8. Now Alice has two equations **(1.11)** and **(1.12)** and two unknowns $\hat{\mathbf{B}}$ and **L**. And she may try to solve them.

The basic idea of the proposed semantic communication protocol is From these considerations we obtain necessary conditions for the semantic relationship for semantic communication systems:

1. Coding operators for semantic relationship must be commutative.
2. Alice must be able to separate semantic relationship **M** and **L**.

Now consider the semantic communication line (Fig. 2)

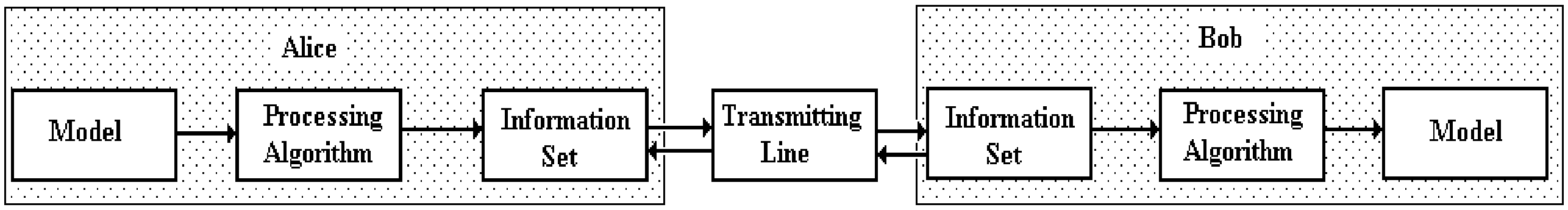

**Fig. 2. Semantic communication line.**

If communication line includes Alice's information set, transmitting line and Bob's information set then communication is possible if Alice and Bob have the same information sets. But if we consider semantic information then it is not necessity that Alice's and Bob's sets were coincident. For semantic communication line it is necessity that Alice's and Bob's models were the same but their information sets may be different.

**1.5. Absolutely Secure Communication over Insecure Channel.**

Let $\mathbf{X_A}$ is the Alice's information set and $\mathbf{X_B}$ is the Bob's information set. Alice has sent a message **M** to Bob. In that case we can write that amount of information which was sent by Alice is

$$\mathbf{I_A = H(X_A) - H(X_A/M)} \qquad \textbf{(1.13)}$$

Bob received the same message **M** and the received amount of information for Bob is

$$\mathbf{I_B = H(X_B) - H(X_B/M)} \qquad \textbf{(1.14)}$$

Eve has intercepted a cipher **C**. And intercepted amount of information is

$$\mathbf{I_E = H(X_E) - H(X_E/C)} \qquad \textbf{(1.15)}$$

Suppose Alice and Bob want to have an absolutely secure communication. It means that $\mathbf{I_E}$ must be equal zero **($\mathbf{I_E = 0}$)**. It is possible if a communication line has loss **L**. In that case

$$\mathbf{I_B = H(X_B) - H(X_B/M) - L} \qquad \textbf{(1.16)}$$

$$\mathbf{I_E = H(X_E) - H(X_E/C) - L} \qquad \textbf{(1.17)}$$

If $\mathbf{I_E} = \mathbf{0}$ then

$$\mathbf{L = H(X_E) - H(X_E/C)} \quad \mathbf{(1.18)}$$

And finally

$$\mathbf{I_B = (H(X_B) - H(X_B/M)) - (H(X_E) - H(X_E/M))} \quad \mathbf{(1.19)}$$

Where the first term $\mathbf{H(X_B) - H(X_B/M)}$ is amount of pure information that would be received without losses and the second term $\mathbf{(H(X_E) - H(X_E/M))}$ describes unavoidable loss.

From the considerations above, it follows that if we have an absolutely secret communication then there is probability of mistakes. If we wish to have unmistakable communication we are not able to have absolutely secure. We are not able to have absolutely secure and absolutely unmistakable communication at the same time.

## 1.6. Absolutely unbreakable cipher.

***Absolutely secure system*** is a cryptosystem that cannot be broken neither in practice nor in theory. No matter :

- how much time an interceptor has (suppose he has unlimited time)
- how many messages were intercepted (suppose he had intercepted all messages)
- what power computer he is using (suppose he has the most powerful computer)
- how much cryptography he knows ( suppose he knows all methods and he also has our software, programs, hardware and knows our cryptography method in details ).

But this definition is not a mathematically strong one. We have no way to determine whether the cryptosystem is absolutely secure or not. In 1949 Claude Shannon introduced [9] the term a ***perfect secure system*** and gave the mathematically strong definition of these systems.

According to the Shannon's theory amount of information **I(M; C)** of the message **M**, that can be revealed if a cipher **C** was intercepted is

$$\mathbf{I(M; C) = H(M) - H(M/C),} \quad \mathbf{(1.20)}$$

And according to Shannon's definition perfect secure system is a system that is

$$\mathbf{I(M; C) = 0,} \quad \mathbf{(1.21)}$$

or

$$\mathbf{H(M) = H(M/C)),} \quad \mathbf{(1.22)}$$

where **H(M)** is amount of information (entropy) in the plaintext. **H(M/C)** is the mutual information of the plaintext and the encrypted text (cipher). It is possible to calculate entropy in the plaintext. But how can **H(M/C)** be calculated in general case if we do not have any information about its key ? We can introduce a universal deciphering algorithm $\Delta_k$ (like a universal Turing machine, recursive function or Markov's algorithm( that can be used for any cipher. If we apply the $\Delta_k$ to actual cipher we get a set of deciphered texts $D = \Delta_k(C)$. After that we define that

$$H(M/C) = H(D) \qquad (1.23)$$

So

$$I_k(M; C) = H(M) - H(\Delta_k(C)), \qquad (1.24)$$

We see that the amount information **I** depends on the parameter **k**, where **k** may be number of algorithmic operations. It allows us to give strong definition for security of public-key system. And give another definition of an absolutely secure system. We will call it an absolutely unbreakable system to differ it from the Shannon's definition of perfect secure system **(1.21)**. So an absolutely unbreakable system is the system that is

$$I_\infty(M; C) = 0. \qquad (1.25)$$

**Note 1.** There is an important difference between the Shannon's definition **(1.21)** and definition **(1.25)**. The Shannon's definition means that **H(M) = H(M/C))**. The definition **(1.25)** means that the task to decrease entropy of its cipher is an algorithmically unsolvable problem.

**Note 2.** $I_\infty(M; C)$ may have negative value. It means that the system has excessive hardness. For example, consider one-time pad method. If the part of one-time pad $I_p$ was intercepted. It means that $I_p$ information can be revealed. If we use a semantic cipher and $I_\infty(M; C)$ has negative value, then the cipher will be absolutely unbreakable even the part of keys was intercepted or cryptographer made mistakes.

## 1.7. Processing semantic information.

To process semantic information we have to find an unsolvable problem **U** that can be used to accomplish an universal set of logical gates. A set of logic gates is said to be universal if any feasible computation can be accomplished in a circuit comprising only gates of the type found in that set. Although there are sets that have a single gate (alternative and joint denial) it is more convenience use AND and NOT gates. The semantic relationship **U** must be distribution semantic relationship. So

$$\hat{A}(M_1)\otimes(\hat{A}M_2) = \hat{A}(M_1\otimes M_2), \qquad (1.26)$$

Where $M_1$ and $M_2$ are initial sets.

### 1.8. Explore a researched object.

The concept of semantic information opens a new way to explore a researched object. The only way that it is applied in science is to get information about the researched object using information sets of properties. But the semantic information concept allows us to arrange research like semantic communications between two partners. This method does not mean that both partners must be alive and intelligent ones. We can arrange semantic information interaction with insensate objects also or research another information system even if it does not know or use an appropriate algorithm.

There are an infinite numbers of unsolvable problems in mathematics (more strictly : the number of these problem is a enumerable set). Any unsolvable problem has unique properties and can be used to make an absolutely unbreakable information system.

## 2. Semantic Theory of Information Applied to Cryptology.

### 2.1. Cryptographic concepts.

Now we explain our idea about the absolute secret communication using visual images. Suppose Alice wishes to send a secret message to Bob. What should she do?

#### 2.1.1. Secret-key concept.

The secret-key concept is the traditional concept in cryptography. In secret-key communication we use the same key for encryption and decryption. If Alice wants to send a message to Bob, she encrypts it using her secret key in the forward direction. When Bob receives this message he decrypts it using the
same key in the inverse direction.

We can illustrate this using a safe analogy (see Fig. 3)
If Alice wants to send a secret message to Bob, she must go through the following steps :

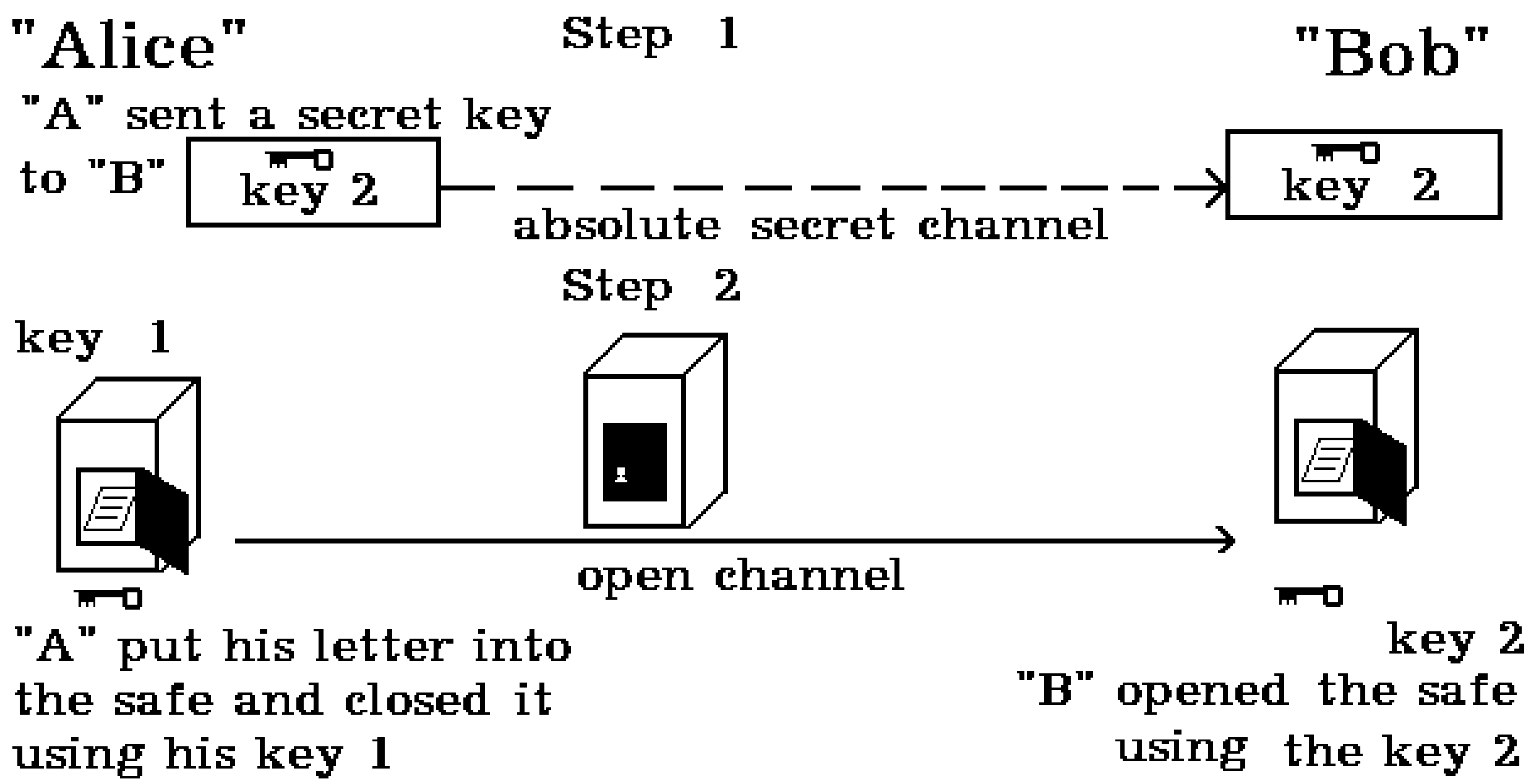

**Fig. 3. Secret-key concept**

1. Alice makes a safe and two keys for it.
2. Alice sends one key to Bob through a secret channel.
3. Alice writes her message, puts it into the safe, closes it and sends this safe to Bob
4. When Bob receives this safe, he opens it using the second key and gets the mess from Alice.

**2.1.2. Public-key concept.**

If Alice uses a method of the Public-Key concept she will do (Fig. 4) :

**1.** She sends an open letter to Bob. She asks him to send a special safe. This safe has a slot. Anybody can put a letter into the safe but it is necessary to have a key to open this safe. Bob only has the key.

**2.** Bob sends this safe to Alice.

**3.** Alice puts her letter into the safe and sends the safe back to Bob.

**4.** Bob opens the safe using his key and takes out Alice's letter.

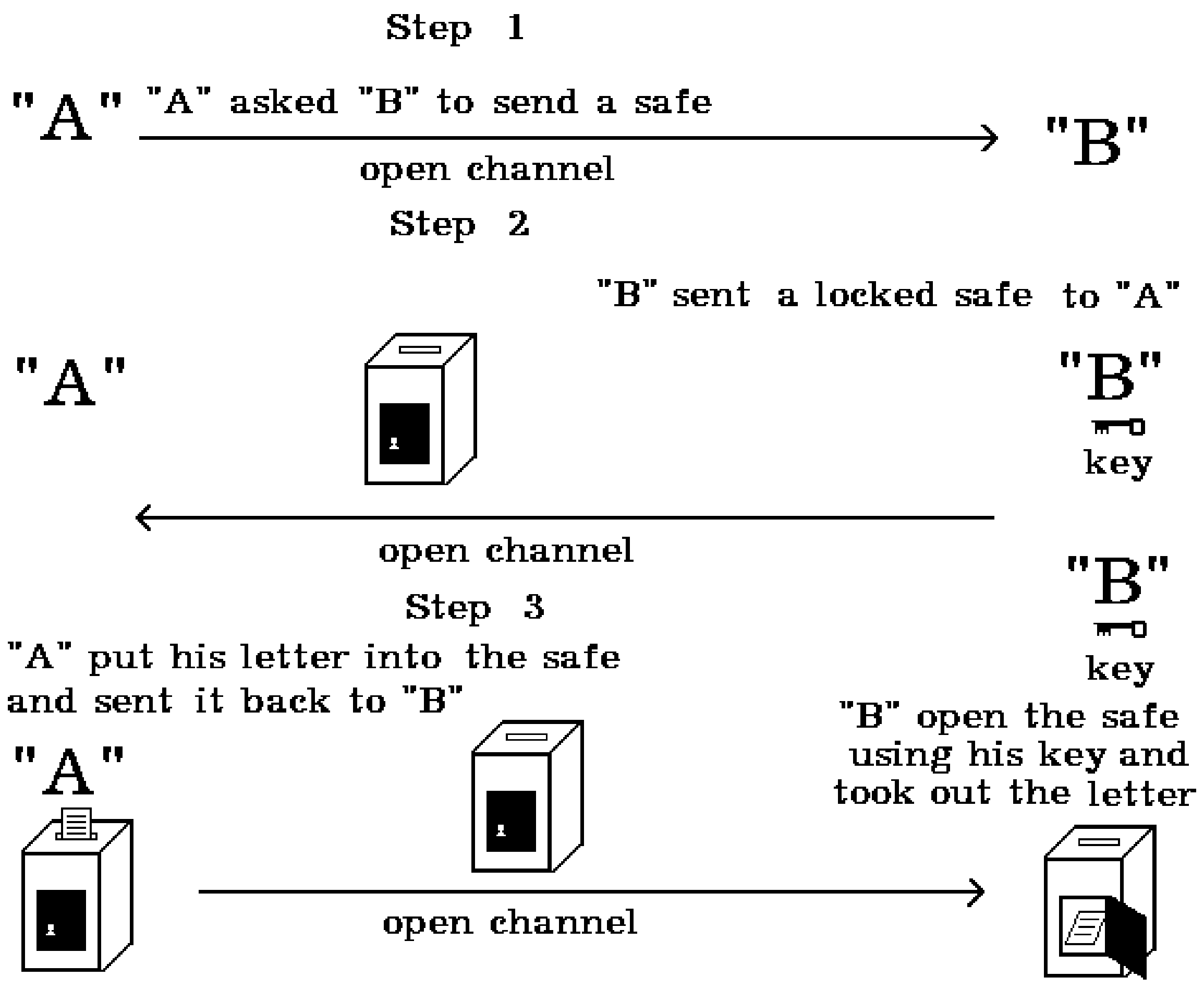

**Fig. 4. Public-key concept.**

### 2.1.3. Double-ciphering concept.

Fig. 1 demonstrates the basic idea of the Double-ciphering conception. Let there be two correspondents - Alice and Bob and the eavesdropper (a passive attacker) Eve. Let Alice sends a secret message to Bob over an insecure channel. She will perform the following steps:

**Step 1**. Alice makes a safe with two doors.

**Step 2.** Alice installs her lock on the first door and keeps the key.

**Step 3.** Alice sends the empty safe with still non-locked second door to Bob with the request to Bob to install his lock on the second door and lock both doors.

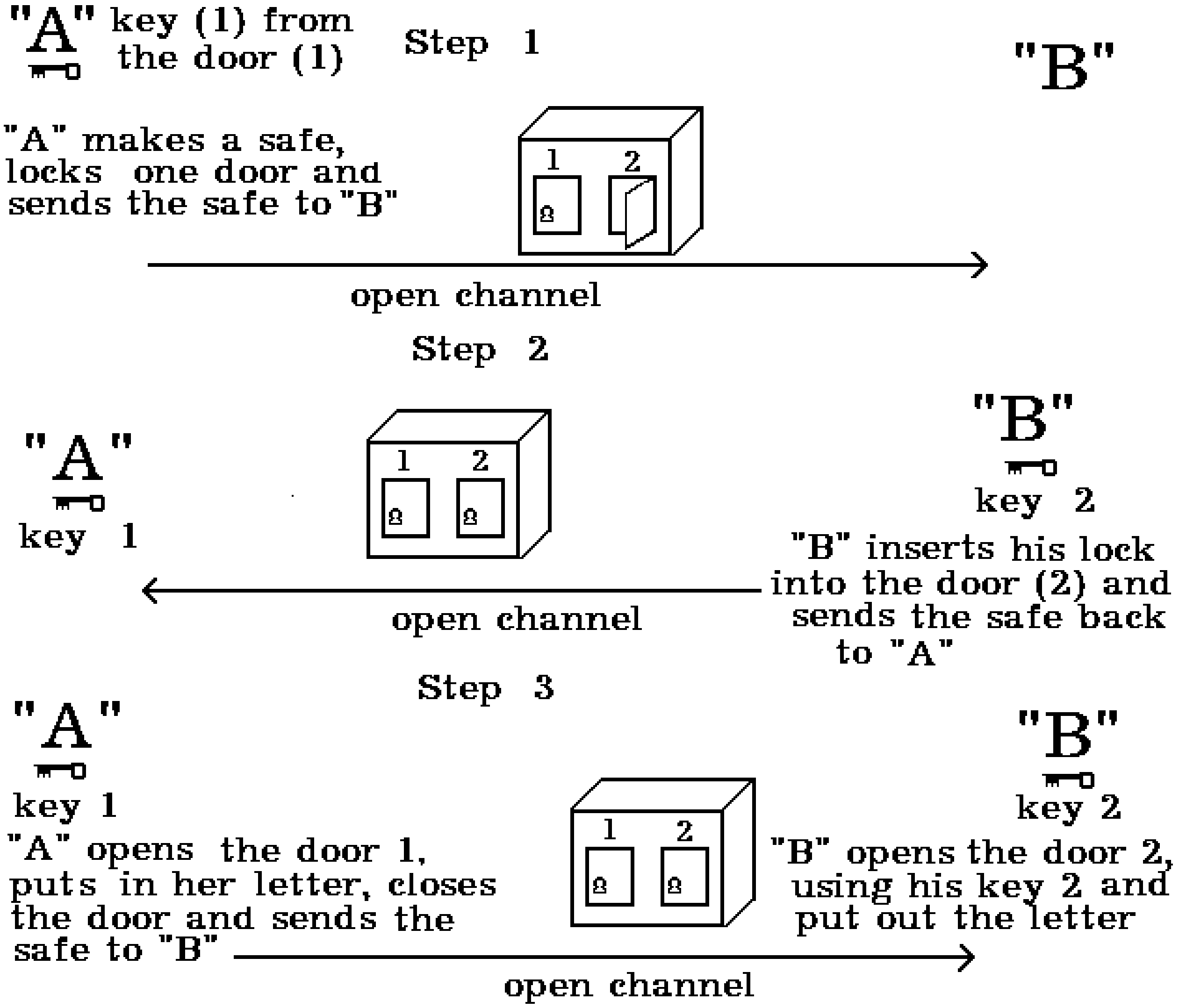

**Fig. 5. Double-ciphering concept**

**Step 4.** Bob installs his lock.
**Step 5.** Bob locks both doors.
**Step 6.** Bob sends the safe back to Alice and keeps his key.
**Step 7.** Alice receives the safe with both doors locked. She opens the first door with her key.
**Step 8.** Alice puts the message into the safe via door 1, locks the door 1 and sends the safe to Bob while keeping her key.
**Step 9.** Bob receives the safe with both doors locked. He opens door 2 with his key.

Since the keys have never left the hands of Alice and Bob, Eve having intercepted the safe, will never be able to open it.

### 2.2. Definitive cryptographic equation.

Every cryptography concept can be described a mathematical equation. This mathematical equation we will call as a definitive equation.

#### 2.2.1. Secret-Key concept.

Let **M** represent a message. Alice perform an operation for coding it. That implies that Alice use a mathematical operator **E**. The result is a cipher **C**. There is an inverse operator $\mathbf{E^{-1}}$ We can calculate the inverse operator if we know **E**. So we can write

$$\mathbf{C = E*M} \tag{2.1}$$

$$\mathbf{E^{-1}*E = I} \tag{2.2}$$

$$\mathbf{M = E^{-1}*C} \tag{2.3}$$

where **I** is identity operator. Any cryptography method that belongs to the secret-key concept can be describe by this equation. So equations **(2.1)-(2.3)** is definitive equation for the secret-key concept.

#### 2.2.2. Double-ciphering concept.

At first, we consider a definitive equation for Double-Key concept. Public-Key concept we will consider later. If we wish to get a definitive equation, we should go through all steps of the Double-Key concept idea.

**1.** Alice constructs a mathematical object **S**(safe). It will represent a safe with two open doors.

**2.** Alice locks the first door. This operation can be described by an operator **A** (Alice). So the **A** operates on the **S**. After that, Alice sends the object **A*S** (the safe with one locked door) to Bob.

**3.** Bob puts his lock on the second door and puts his letter into the safe. It means that Bob have added an object **L** (lock). He has **A*S + L** (The sign "+" has symbolic meaning of addition).

**4.** Bob locked his door. It means that he has performed an operation **B** (Bob) on the object **A*S + L**. He sends the object **B*(A*S + L)** to Alice.

**5.** Alice opens her door. The inverse operator **A-1** operates on the object **B*(A*S + L)**. After that she will get cipher **C**. Alice will be able to decode it and read Bob's letter. We have described all steps of the Double Key concept. So we can describe the Double-Key concept by the equation :

$$\mathbf{C = A^{-1}*B*(A*S + L) = A^{-1}*B*A*S + A^{-1}*B*L} \tag{2.4}$$

Using the definitive equation we can understand why Eve is unable to decrypt Alice's and Bob's messages and how Alice and Bob can read their letters.
The first time (Alice -> Bob) Eve has intercepted cipher

$$\mathbf{C_1 = A*S} \quad \mathbf{(2.5)}$$

Eve has one equation **(2.5)** and two unknowns **A** and **S**. So she cannot solve this equation.
The second time (Alice <- Bob) Eve have intercepted cipher

$$\mathbf{C_2 = B*(A*S + L)} \quad \mathbf{(2.6)}$$

She knows **A*S** but she does not know **B** and **L**. So she has one equation and two unknowns again. So she are not able to solve the equation. Alice decrypts **B*(A*S + L)** using her key (the inverse operator $\mathbf{A^{-1}}$) for decryption. We can say that she has put away her operator **A** for encryption (we suppose **A*B = B*A**) and she gets :

$$\mathbf{C_3 = A^{-1}*B*(A*S + L) = B*S + A^{-1}*B*L} \quad \mathbf{(2.7)}$$

Alice has one equation and two unknowns **B** and **L** too. But she can try to separate **B*S** from **A-1*B*L**. In this case Alice will get

$$\mathbf{C_4 = B*S} \quad \mathbf{(2.8)}$$

Alice knows $\mathbf{C_4}$ and **S**. She can try to solve the equation, finds Bob's operator **B**, and use it for encrypting her message and read Bob's message **L**. Therefore **S** should have a special mark or another property for separation. This mark should be secret for Eve. Bob's operator **B** should not change it. Eve is unable to see the **S** mark because Eve knows only **A*S**. We can say that the operator **A** "hides" the secret mark. But there is a question. How can we mark **S** ? The mark should be absolute secret if we want to have an absolute unbreakable cipher. Eve has intercepted **A*S** and $\mathbf{A^{-1}*B*L}$. So she is able to analyze them and find all their properties. How can these objects have absolute secret properties. We will consider this question in the next chapter. Now we should write a definitive equation for Public-Key concept.

### 2.2.3. Public-Key concept

We have seen the idea of Double-Key concept. Now let see how Public-Key concept can be described using double-key definitive equation. We saw that Bob can send his letter to Alice when he send back Alice's safe. If Alice and Bob use Public-Key concept Bob does not add anything into Alice's safe. So we will get the definitive equation for Public-key concept if suppose that

$$\mathbf{L = 0.} \quad \mathbf{(2.9)}$$

In this case we have

$$C_2 = B*(A*S) \qquad (2.10)$$

What does it means ?
Suppose **(A*S)** does not have an inverse element. Eve knows **A*S** but she cannot figure out **B** as

$$B = C_2*(A*S)^{-1} \qquad (2.11)$$

But **A** has an inverse operator. Eve cannot figure out it because she intercepted only **A*S**. So Eve has one equation and two unknowns **A** and **S**. But Alice knows **A**. And **A** has an inverse operator. So Alice can figure out $A^{-1}$ and write

$$C_4 = B*S \qquad (2.12)$$

Idea of the Public-Key concept is that **S** is very simple. But **A*S** is very difficult. So equation **(2.12)** is more possible for solution than equation **(2.11)**. **A*S** is a public key and **A** is a private key

### 2.2.4. Why we need definitive equations ?

Cryptographic science does not have strong definitions for its cryptographic concepts. Session key and updating key methods look like Double-Key concept method. Definitive equations allow us to differentiate methods.
For example: session key protocol may be describe as updating key protocol. We see that Secret-Key concept is a special case of Double-Key concept. If we get **A = B** and **S = 0** we will get that Alice receive a cipher from Bob It is the same equation. We have seen that both the Secret-Key concept and the Public-Key concept are special cases of the Double-Key concept :

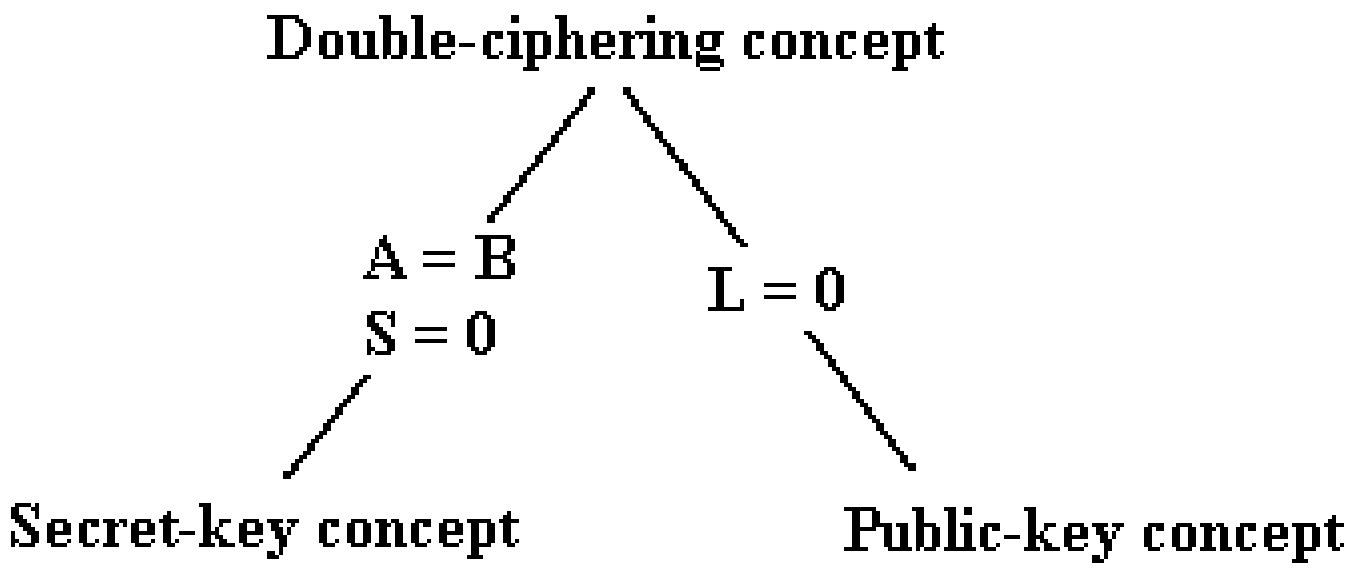

**Fig. 6. Cryptographic concepts**

But the Secret-Key concept is not converted to Public-Key concept. Therefore there is not misconception between Secret-Key and Public-Key methods. But it is easy to make a mistake between Public-Key concept and Double-Key concepts.

### 2.3. Double-ciphering communication Line.

#### 2.3.1. Equations.

The simplest semantic relationship is that one element is produced from another elements. So there are **n** independent elements $\{O_1, \ldots, O_n\}$. And

$$O_{n+1} = A(O_1, \ldots, O_n), \qquad (2.13)$$

where **A** is a semantic algorithm. We saw above (see 1.4) that it is necessary to have two commutative algorithms **A** and **B** for semantic communication. Let **B** is another algorithm.

$$B(A(O_1, \ldots, O_n)) = A(B(O_1), \ldots, B(O_n)) \qquad (2.14)$$

In that case we can make a semantic communication line in the following way :

We have two necessary condition for double-key method realization:
**1.** Two strong one-way functions **F** and **T** (or in general algorithms). The function (algorithm) **F** has **n** arguments and the function (algorithm) **T** has one argument. So

$$F = F(O_1, \ldots, O_n) \qquad (2.15)$$
$$T = T(O_1) \qquad (2.16)$$

A strong one-way functions means that :
**First.** We can not calculate the inverse functions. The equations

$$C_1 = T(X) \qquad (2.17)$$
$$C_2 = F(X_1, \ldots, X_n) \qquad (2.18)$$

(where $X, X_1, \ldots, X_n$ are unknowns and $C_1$ and $C_2$ are arbitrary constants) are insolvable.
**Second.** For arbitrary constant $C_1$ and $C_2$. We cannot decide whether they do or do not have a solution.

**2.** The function **F** and **T** must be commutative to each other. It means that

$$T(F(O_1, \ldots, O_n)) = F(T(O_1), \ldots, T(O_n)) \qquad (2.19)$$

#### 2.3.2. The cryptographic process. The first level.

**The first Alice session.**

**Step 1. Alice makes a safe.**

Alice makes an initial cryptographic framework (safe). The cryptographic framework consists of **n** mathematical objects $\{O_1, O_2, \ldots, O_n\}$.

**Step 2. Alice installs her lock on the first door.**
There is a strong one-way function

$$O_{n+1} = F(O_1, O_2, \ldots, O_n) \qquad (2.20)$$

The value of the function **F** will change if transposition of its arguments is made. The function **F** is known only to Alice. It is her key. Alice calculates $O_{n+1}$.

**Step 3. Alice sends the empty safe to Bob.**
Alice sends the broaden cryptographic framework $\{O_1, O_2, \ldots, O_n, O_{n+1}\}$ to Bob.

**The first Bob session.**

**Step 4. Bob installs his lock.**
Bob transforms the cryptographic framework objects $\{O_1,\ldots,O_n, O_{n+1}\}$ using a secret function **T**. The **T** is known to Bob only. It is his key. The **T** is a strong one-way function too. Functions **T** and **F** are commutative to each other

$$T(F(O_1, O_2, \ldots, O_n)) = F(T(O_1), T(O_2), \ldots, T(O_n)) \qquad (2.21)$$

**Step 5. Bob locks both doors.**
Bob makes a permutation of the framework objects, deletes originally received from Alice objects $\{O_1,\ldots,O_n, O_{n+1}\}$ and leaves the following set only:

$$\{ T(O_{k_1}), T(O_{k_2}), \ldots , T(O_{k_{(n+1)}})\} \qquad (2.22)$$

**Step 6. Bob sends the safe to Alice.**
Bob sends the permutated cryptographic framework (2.22) back to Alice.

**The second Alice session.**

**Step 7. Alice opens the safe.**
Alice re-builds the cryptographic framework back to the original order by looking through all possible transpositions and using the **F** function for verification. Thus, Alice can compute Bob's permutation. Now she uses it as the cipher key for making the secret message to be sent. This operation accomplishes the scope of level one.

**The second Bob session.**

**Step 8. Bob opens the safe.** Bob uses his key to decode Alice's message.

### 2.3.3. The cryptographic process. The second level.

Regardless of the fact that **F** and T functions are absolutely strong one-way functions the first cryptographic system is not absolutely secret because it is vulnerable to the brute-force attack. This brute-force attack can be withstood by adding the second level of the cryptographic system.

**Step 1.**

Alice converts her initial message (plaintext) to a binary format. The converted message is a string of **0s** or **1s**. Alice and Bob make an agreement about the conversion rule using an open channel before the secret communication session begins.

| | Conversion Rule | | |
|---|---|---|---|
| Initial message | ----------------→ | Binary format | **(2.23)** |

***************************************************************************

**Example.** Suppose an initial message is a word "No". If we use ASCII conversion the converted message becomes "10011101101111".

| | |
|---|---|
| Initial message | No |
| Conversion rule: | ASCII conversion |
| Binary initial message | 0100111001101111 |

***************************************************************************

**Step 2**.

Alice makes cryptographic binary words for every digit of the initial binary message. A cryptographic binary word is a string of **n** **($n \geq 2$)** bits. Therefore, there are $2^n$ different cryptographic words. Alice and Bob make arrangement through open channel about which cryptographic words correspond to **0** and **1**. The words made out of **11...1** becomes a decoy word as it carries no information essential to the secret message. Alice randomly selects cryptographic words for every digit of her binary initial message. If she encounters the word **11..1** she keeps the word but repeats the selection for this digit again. This can be illustrated by the following example.

***************************************************************************

**Example.**

Suppose Alice and Bob have made an arrangement that if a cryptographic word has even ones then it presents **0**. If it has odd ones then it presents **1**. Let cryptographic words have 4 digits (**n = 4**). In that case the words

**0011, 0101, 0110, 1001, 1010, 1100, 0000** correspond to **0** **(2.24)**

And the words

**0001, 0010, 0100, 1000, 0111, 1011, 1101, 1110** correspond to **1** **(2.25)**

The word

**1111** is the decoy word **(2.26)**

***************************************************************************

**Step 3.**

Alice transmits every digit of the binary cryptographic words to Bob using the level one of the cryptographic system as given below

1. If she transmits **1** she makes an additional object $\mathbf{O_{n+1}}$ using function **F**. If she transmits **0** she does not use function **F** and makes the object $\mathbf{O_{n+1}}$ by random. In that case cryptographic framework has not function **F**.
2. Bob makes transposition and sends the cryptographic framework back to Alice.
3. If Alice transmits **1** she can calculate Bob's transposition and sends to Bob the correct transposition number. If she transmits **0** she cannot calculate Bob's transposition and sends a random transposition number. Alice and Bob have an agreement that if Alice sends the correct transposition number it means **1**. If Alice sends the incorrect transposition number it means **0**.

Security of the first level cryptographic system is based on the secret function **F** . If Eve wants to break the cipher she has to find the **F**. Eve knows exactly that **F** exists. Therefore she makes a brute-force attack. It is an algorithmically defined process.

Security of the second level cryptographic system is based on the knowledge that the function **F** exists or not. Eve does not know if **F** exists. To break the cipher Eve has to find **F** or prove that the function **F** does not exist. However, it can be demonstrated that Alice will be able to choose from those functions that are algorithmically unsolvable mathematical tasks. It means that Eve will not be able to prove existence of **F**.

The Secret-Key conception uses one to one function. The Public-Key conception uses one-way function with drop door. The proposed Double-Key conception allow use one-way function without drop door. It allows for the use of a strong one-way function and makes an absolutely unbreakable cipher.

### 2.4. Real Absolutely Secure Communication System.

We can build a real absolutely secure communication system if we use Diophantine functions as the **T** and **F** functions. Solution of Diophantine equations is an algorithmic unsolvable task [7]. There are a lot of commutative Diophantine equations and many of them can be used for building real absolutely secure communication systems.

## 3. Advantages of Double Key concept and its applications.

### 3.1. Advantages of Double key concept.

Double Key concept has the following advantages :

1. Alice's key and Bob's key are independent of each other.
2. Double key concept uses one-way functions without drop door.
3. Double key methods allow to work in infinite fields.
4. There are unlimited number of Bob's key which match with Alice's key. And the contrary is right too.
5. Secret key concept and Public key concept are special cases of Double key concept.
6. Semantic information may "go through" algorithms untouched.

### 3.2. Applications.

Advantages 1-3 allows to make absolutely secure communication systems.

Advantages 4 –5 allows to make a new class of applications.

We think the most important advantage is number 5 and 6. It allows to develop a new concept and methods for deciphering. These method will be the most powerful method. In our opinion any cryptographic method which ciphering algorithm is determined (any public key method is determined algorithm) can be broken for polynomial number of steps.

.

**3.3. Demonstration programs.**

We have made demonstration programs for

1. Secure communication between two correspondents.
2. Protection password file for firewall.
3. Protection Credit card.

## References.